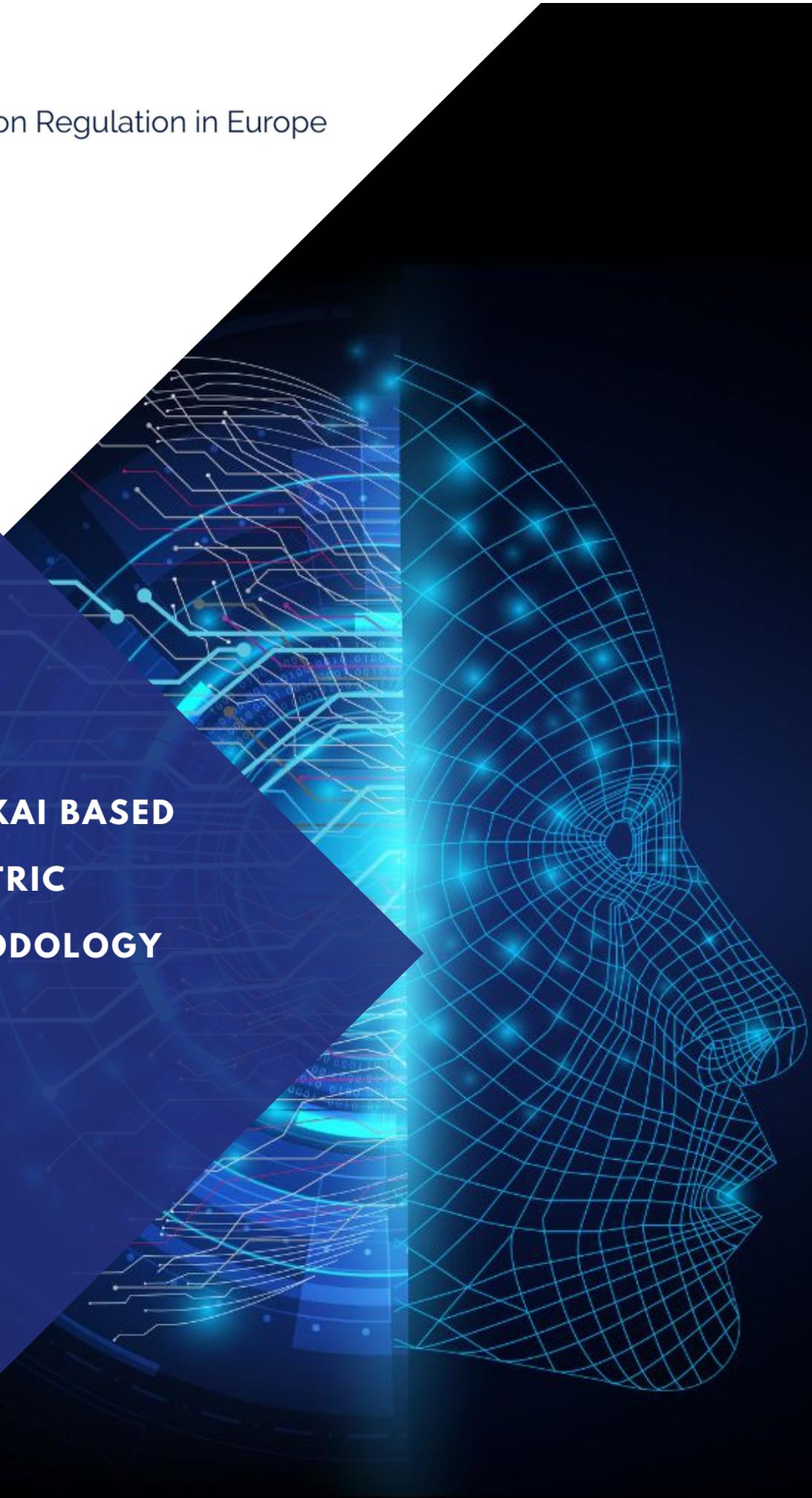

cerre | Centre on Regulation in Europe

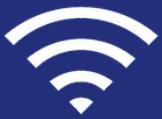

# MEANINGFUL XAI BASED ON USER-CENTRIC DESIGN METHODOLOGY

REPORT

*July 2023*

Winston Maxwell
Bruno Dumas

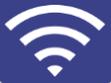

# MEANINGFUL XAI BASED ON USER-CENTRIC DESIGN METHODOLOGY

## COMBINING LEGAL AND HUMAN-COMPUTER INTERACTION (HCI) APPROACHES TO ACHIEVE MEANINGFUL ALGORITHMIC EXPLAINABILITY

### JULY 10, 2023

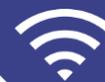

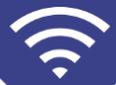

# Table of Contents

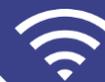

## ABOUT CERRE

Providing top quality studies and dissemination activities, the Centre on Regulation in Europe (CERRE) promotes robust and consistent regulation in Europe's network and digital industries. CERRE's members are regulatory authorities and operators in those industries as well as universities.

CERREs added value is based on:

- ▪ its original, multidisciplinary and cross-sector approach;
- ▪ the widely acknowledged academic credentials and policy experience of its team and associated staff members;
- ▪ its scientific independence and impartiality;
- ▪ the direct relevance and timeliness of its contributions to the policy and regulatory development process applicable to network industries and the markets for their services.

CERREs activities include contributions to the development of norms, standards and policy recommendations related to the regulation of service providers, to the specification of market rules and to improvements in the management of infrastructure in a changing political, economic, technological and social environment. CERREs work also aims at clarifying the respective roles of market operators, governments and regulatory authorities, as well as at strengthening the expertise of the latter, since in many Member States, regulators are part of a relatively recent profession.

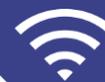

# ABOUT THE AUTHORS

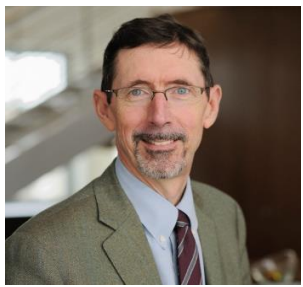

**Winston Maxwell** is Director of the Law & Digital Technology Studies department at Télécom Paris – Institut Polytechnique de Paris, where he teaches and writes on subjects related to the regulation of data, AI and telecommunications.

He previously had a career in private practice as a partner of the international law firm Hogan Lovells.

Winston completed his law degree (JD) at Cornell, his PhD in economics at Télécom Paris, and his HDR (Habilitation à Diriger des Recherches) at the University of Paris Panthéon Sorbonne. His research focuses on the regulation of AI, and in particular human control over algorithmic systems, explainability and bias. Winston co-ordinates the "Operational AI Ethics" program at Telecom Paris, which includes AI Ethics teaching at Institut Polytechnique de Paris. In addition to being a research fellow at CERRE, Winston is a member of the Data and AI Ethics Council of Orange, the Scientific Advisory Board of ARCOM, the Ethics Board of the Paris Institute of Advanced Studies. He also contributes to standardisation activities on trustworthy AI within ISO/IEC JTC21/WG 4.

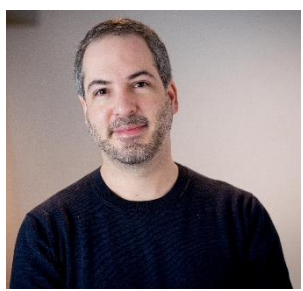

**Bruno Dumas** obtained his PhD in computer science in 2010 at the University of Fribourg, in Switzerland. His thesis focused on the creation of multimodal interfaces, along three axes: software architectures for multimodal interfaces, multimodal interaction modelling, and multimodal fusion algorithms.

He then worked for four years as a post-doc researcher at the Vrije Universiteit Brussel, in Belgium, where he deepened his knowledge of cross-media interactive systems. He is now a professor at the University of Namur, where he leads the EXUI research group, and is the co-president of the NADI research institute. His main research domains revolve around human-computer interaction, multimodal interfaces, augmented/mixed reality, as well as how the evolution of IT impacts users in their daily life and our society.

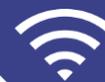

# ABSTRACT

This report explores the concept of explainability in AI-based systems, distinguishing between "local" and "global" explanations. "Local" explanations refer to specific algorithmic outputs in their operational context, while "global" explanations encompass the system as a whole. The need to tailor explanations to users and tasks is emphasised, acknowledging that explanations are not universal solutions and can have unintended consequences. Two use cases illustrate the application of explainability techniques: an educational recommender system, and explainable AI for scientific discoveries. The report discusses the subjective nature of meaningfulness in explanations and proposes cognitive metrics for its evaluation. It concludes by providing recommendations, including the inclusion of "local" explainability guidelines in the EU AI proposal, the adoption of a user-centric design methodology, and the harmonisation of explainable AI requirements across different EU legislation and case law.

Overall, this report delves into the framework and use cases surrounding explainability in AI-based systems, emphasising the need for "local" and "global" explanations, and ensuring they are tailored toward users of AI-based systems and their tasks.

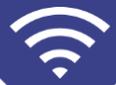

# EXECUTIVE SUMMARY

This report explores the concept and role of explainability in AI-based systems. It considers explainability through the framework of 'local' and 'global' explanations: local explanations refer to a particular algorithmic output, and are sensitive to their operational context, while global explanations refer to the functioning of the system as a whole, not to particular algorithmic outputs.

The report considers questions about the characteristics of end users of AI-systems, their tasks, and the information they need to effectively use the systems. It emphasises that explanations are not a universal solution and can sometimes have unintended consequences, such as making users complacent or exacerbating their cognitive biases.

The report presents two use cases to illustrate the application of explainability techniques. The first use case involves an educational recommender system that provides learners with explanations for recommended learning material based on their performance in being examined on it. This system includes visualisations and performance comparisons to help learners identify areas for improvement. The second use case focuses on explainable AI for scientific discoveries, where explanations play a role in aiding scientists to understand the output results or models of machine learning algorithms.

The report acknowledges that meaningfulness is a key aspect of explanations and discusses its interpretation in the context of the European Union General Data Protection Regulation ('GDPR'). It highlights that meaningfulness is subjective and depends on various factors such as user characteristics, context, and task. The concept of meaningfulness is further explored through dimensions of coherence, purpose, and significance, which are described as levels leading to meaningful interactions.

The report considers cognitive metrics to measure and assess the meaningfulness of explanations, such as 'explanation goodness', user satisfaction, user trust, and user understanding. It states that qualitative and quantitative evaluation techniques involving user feedback, questionnaires, and interviews can be used to assess meaningfulness.

Overall, the report highlights the importance of tailoring explanations to the user and task at hand, while considering the potential impact and challenges in measuring meaningfulness. It concludes by stating that the field of explainability techniques for AI-based systems is still developing, and makes the following recommendations:

**For the EU AI proposal:**
1. **Specific guidance on 'local explainability'.** The absence of guidelines for local explainability in the AI proposal is seen as unfortunate, considering that other EU texts already require forms of local explainability. The AI proposal should acknowledge the need for local explainability in the risk analysis and management system of the AI provider, and which should be adapted as necessary by the user of the AI system. This will empower the European Commission to develop guidelines on when and why AI explanations are necessary, and how they should be "meaningful" based on their purpose and audience. The creation of such guidelines could harmonise the 'explainability' requirements across different sources of EU law.

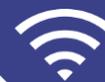

**2. A user-centric design methodology** as a way of measuring whether a given explanation is "meaningful" in light of its audience and purpose.

**3. The EU Charter could underpin the approach to explainability in AI-based systems** and provide a foundation for seeking a harmonised methodology.

**General Complementary Recommendations:**

**1. European standardisation bodies could develop a framework for meaningful explanations** in the context of harmonised standards.

**2. For the European Centre for Algorithmic Transparency,** established by the Commission to deal with algorithmic transparency in the context of the DSA, **to develop a user-centric design methodology** of the kind proposed in this report.

**3. The European Commission should pursue:**
- **harmonisation of multiple explainable AI requirements** in different EU legislation and case law; and
- **encouragement of user-centric design methodology** to achieve 'meaningful' explanations.

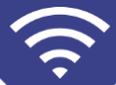

# 1. INTRODUCTION

Although the term "explainability" of Artificial Intelligence ('AI') never appears as such in European Union ('EU') legislation, the requirement of transparent and understandable algorithms ('XAI' - Explainable Artificial Intelligence) is present in many EU regulations, directives, legislative proposals, and case law, including the proposed AI Act.[1] These legal instruments address multiple objectives of XAI, including protecting individual autonomy, ensuring effective human oversight, ensuring contestability of algorithmic decisions, preserving due process rights, effective audit and traceability of AI systems, and addressing information asymmetries and imbalances of economic power. Most of these objectives focus either on the user of the AI system – the person overseeing the algorithmic decision system – or on the person who is affected by an algorithmic decision. The explanation must be meaningful or useful for the person to whom the explanation is given, the "recipient" of the explanation, in light of the objective targeted by the legislation.

To date, most research on XAI focuses on the needs of data scientists, who use explainability to improve their models. However, a growing literature in the domain of human-computer interactions ('HCI') is studying XAI from the standpoint of user behaviour and needs. This user-centric approach converges with the focus of regulation, which is also to maximise the utility of explanations for users and individuals affected by decisions. Explainability is not a principle to be protected in its own right, but a means to enhance human control and human rights. Consequently, XAI must be developed, and its success measured, with the final objectives of XAI in mind.

Experience in applying the GDPR[2] has shown that transparency requirements can turn into formal box-ticking exercises, disconnected from human cognitive realities. Many data processing operations involve multiple steps, purposes and actors, and yet information on data processing must be provided in concise, easily accessible and understandable form, using plain language.[3] Simplification is needed to make information understandable, yet simplification comes at the cost of omitting information, some of which may be important, at least to some users. The balance between simplification, on the one hand, and completeness, on the other, has led to considerable legal uncertainty on what disclosures are necessary to ensure that users are sufficiently informed.

The trade-offs between completeness and understandability will arise in the context of XAI, whose purpose is to explain a system or a particular algorithmic output. Explanations of a particular algorithmic output, so-called "local" explanations, are sensitive to their operational context. The time available to the user, the user's technical expertise, and the stakes involved in the individual decision will affect whether a local explanation is needed, what information should be conveyed in the local explanation, its level of simplification, and its form. A trade-off will always exist between ease of understanding, on the one hand, and accuracy and completeness, on the other.

---

[1] European Commission, *proposal for a regulation of the European parliament and of the council laying down harmonised rules on artificial intelligence*, COM(2021) 206 final, 24 April 2021 (Proposed European AI Act).

[2] Regulation (EU) 2016/679 of the European Parliament and of the Council of 27 April 2016 on the protection of natural persons with regard to the processing of personal data and on the free movement of such data, and repealing Directive 95/46/EC (GDPR).

[3] Ibid., recital 58.

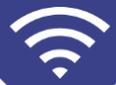

The proposed European AI Act's provisions on transparency impose technical documentation and user notices that explain how the system functions as a whole.[4] These requirements relate to so-called "global" explanations, because they relate to the functioning of the system as a whole, and not to a particular algorithmic output. The proposed AI Act contains no specific requirement on "local" explanations to understand individual outputs, even though we know "local" explanations of this kind will be essential for effective human oversight, safety, and respect for fundamental rights. The regulatory approach to XAI is further confused by the existence of XAI-related requirements in other EU legislation and case law, and the use of different terms to describe these requirements. As we will see in Section 3, the Platform Workers' Directive requires giving "reasons" for individual decisions; the Digital Services Act ('DSA') and Platform to Business regulation require explanation of the "main parameters"; Court of Justice of the European Union ('CJEU') case law requires understanding of "how criteria and programs work"; and the GDPR requires "meaningful information about the logic involved".

In light of the above, this paper attempts to do four things.

**First**, we take stock of XAI-related requirements appearing in various EU directives, regulations, guidelines, and CJEU case law. This analysis of existing requirements will permit us to have a clearer vision of the purposes, the "why", of XAI, which we separate into five categories: contestability, empowerment/redressing information asymmetries, control over system performance, evaluation of algorithmic decisions, and public administration transparency. The analysis of legal requirements also permits us to create four categories of recipients for explainability: data science teams; human operators of the system; persons affected by algorithmic decisions, and regulators/judges/auditors. Lastly, we identify four main operational contexts for explainability: XAI for the upstream design and testing phase; XAI for human-on-the-loop control; XAI for human-in-the-loop control; and XAI for ex-post challenges and investigations.

**Second**, we will present user-centered design methodology, which takes the purposes, the recipients and the operational context into account in order to develop optimal XAI solutions.

**Third**, we will suggest a methodology to permit suppliers and users of high-risk AI applications to propose local XAI solutions that are effective in the sense of being "meaningful", for example, useful in light of the operational, safety and fundamental rights contexts. The process used to develop these "meaningful" XAI solutions will be based on user-centric design principles examined in the second part.

**Fourth**, we will suggest that the Commission issue guidelines to provide a harmonised approach to defining "meaningful" explanations based on the purposes, audiences and operational contexts of AI systems. These guidelines would apply to the AI Act, but also to the other EU texts requiring explanations for algorithmic systems and results.

The remainder of this paper is organised as follows. Section 2 will present the main categories of AI and main approaches to explainability. Section 3 will present an inventory of XAI requirements imposed by laws and draw conclusions on the purposes, audiences and operational contexts of XAI.

---

[4] Proposed European AI Act, Article 13 and Annexe IV.

Section 4 will present user-centric design principles, and how they should be applied to ensure that XAI is meaningful. Section 5 will propose a methodology for developing meaningful XAI solutions based on user-centric design principles and the risk assessment envisaged in the future AI Act. Section 6 will conclude with concrete recommendations for clarifying the legal requirements for XAI, in particular in the proposed European AI Act.

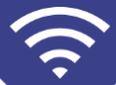

# 2. AN OVERVIEW OF XAI

## 2.1 Symbolic AI Versus Machine Learning

The field of artificial intelligence started quite early in the existence of computer science. Marvin Minsky and John McCarthy coined the word in 1956, during a workshop which aimed at creating a new research area aimed at building machines able to simulate human intelligence.[5] Since then, the term "artificial intelligence" has been applied to a broad number of different computer science fields, few of which are aligned with the original goal of the 1956 workshop. Among these different fields, one may note knowledge representation, computer vision, planning or machine learning. This last field is now the most well-known of all. Dozens of different machine learning techniques exist today. De Spiegeleire et al. 2017[6] present a good introduction to the spectrum of artificial intelligence techniques. Since 2012, deep learning techniques and their applications have received the lion's share of media attention. Explainability techniques are a bit more recent than artificial intelligence, however as AI and, in particular, machine learning based systems grew in complexity, the question of interpreting and understanding their behaviour appeared quickly. Swartout et al. 1991 present a good illustration of the application of explainability principles applied to expert systems in the 1980s and 1990s.[7]

Among the dozens of AI and machine learning techniques, some are inherently explainable by their very nature. One may for example mention decision trees, which work by guiding their reader through a series of binary, "if-then", choices. These choices are extracted from a set of training data, and probabilistic values illustrate the most likely path. However, many other techniques, among which some of the more popular ones currently, work like "black boxes". These methods use an algorithm to create a "model" based on a set of training data. The algorithm can be applied to any use case, and for a given use case, a specific set of training data will create a specific model. If the algorithms are generally well known, and their code is frequently publicly available, issues arise when one wants to probe models. As an example, models created for neural networks consist of huge quantities of "weights", or values describing the probability that a transition from one perceptron to another is triggered. Even for very basic examples, these values are meaningless when reviewed raw. This issue quickly compounds when one considers the high level of complexity of current deep neural networks used nowadays for the most advanced applications, including very large foundation models. Trying to make sense of these "black boxes" has been an important research question in the field of computer science for a few decades already, and the accelerating adoption of machine learning techniques has only made the issue more pressing. New techniques appear regularly, however, the parallel added complexity of new machine learning techniques creates a kind of game of catch-up between novel explainability techniques and novel AI-based techniques.

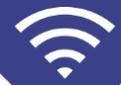

## 2.2 Explainability Involves Trade-Offs

Explanations are not free. In addition to taking up human time and computer resources, they may actually degrade predictive performance, by forcing developers and users to prefer rule-based models instead of machine learning models, even when the latter perform better.[8] In high-stakes situations, imposing explainability will often be well worth the cost, and the explainability/performance trade-off will become less pronounced as data scientists develop new machine learning approaches that also integrate explainability constraints. However, it is fair to say that explainability will always generate a cost which needs to be balanced against the benefits derived from it. The inevitable trade-offs involved in XAI lead us to recommend, in Section 5, that XAI solutions be evaluated in the context of risk assessments for high-risk AI applications.

Where the stakes are high, for example investigating the cause of a serious accident, the benefits of requiring detailed explanations will be very high, because explanations could help determine who is responsible for the accident and help prevent similar accidents in the future. In a situation with lower stakes, for example finding out why Netflix recommended a particular television series, the social benefits of requiring detailed explanations will be low compared to the costs. Each situation warrants its own level of explanation, going from light explanations for algorithmic decisions with little or no consequence to detailed explanations for high-stakes algorithmic decisions.

HLEG (2019) expresses the cost-benefit trade-off in the form of a sliding scale: "*The degree to which explicability is needed is highly dependent on the context and the severity of the consequences if that output is erroneous or otherwise inaccurate*".[9]

As explained in Section 3.2, the CJEU has gone so far as to exclude opaque machine learning from certain applications with high impacts on individual rights, because of the inability of human users to understand the "reasons" for a particular algorithmic output. This could lead governments and companies to sacrifice predictive performance for the sake of explainability, a choice that could create its own costs, particularly in algorithms designed to detect fraud or other criminal activity.

## 2.3 Explainability is "Global" or "Local"

The definitional questions surrounding explainability are complex. Some scholars distinguish between interpretability and explainability, while others distinguish between transparency and explainability.[10] One of the main functional differences relates to whether a person understands what is being explained. A complete and accurate explanation may be given, thereby achieving transparency, but if the explanation is not understood by the recipient, there will be a failure in the interpretability, or

---

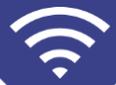

"meaningfulness", of the explanation.[11] Some scholars refer to "interpretable AI" as a form of inherently understandable model, while "explainable AI" refers to post-hoc methods to make decisions of black-box models understandable to humans.[12] Others refer to "explainability" as a representation of an AI system's operation, whereas "interpretability" refers to the meaning of an AI system's output.[13] Instead of entering into this complex definitional debate, we prefer to simplify the definitional question by dividing XAI into two main families, which we (and others) call "global" and "local" explanations.[14] This simplification masks the nuances between different forms of XAI. In particular, we include within the term "global" explainability both black box models, the overall functioning of which is explained through post-hoc XAI, and interpretable models, the functioning of which is inherently understandable without any post-hoc explanation. Our task in this paper is to understand what "meaningful" XAI is. Dividing explanations into "global" and "local" is useful for this task, because it permits us to focus on the purposes and effects of explanations on their intended recipients.

"Global" explanations convey an understanding of the system as a whole; "local" explanations convey an understanding of a particular algorithmic output. These two approaches serve different purposes and rely on different techniques. A global explanation will rely on technical and user documentation to inform a user or auditor about the overall operation of the algorithm, its weaknesses, how it was developed and trained, and its approved use environments. Annex IV of the proposed European AI Act requires extensive information on "global" explainability for high-risk AI systems, much like a user notice accompanying a medicine describes the active molecules, the recommended dosages, the counter-indications, and the side effects. The proposed European AI Act embraces this approach for all high-risk AI applications, by imposing technical documentation, user manuals, and human oversight requirements that ensure global explainability. Other approaches to global explainability rely on so-called "model cards",[15] or "accountability reports".[16] The level of detail included in global explanations must be balanced against protection of security. Giving too much information about the functioning of the algorithm could facilitate adversarial attacks. Similarly, giving too much information, including source code, could interfere with the legitimate protection of trade secrets.[17]

"Local" explanations concern individual algorithmic outputs, such as a risk score. Local explainability refers to the ability of a user, citizen, or data scientist, to understand a specific algorithmic result, for example, by being able to answer the following types of questions in their context: '*Why did the algorithm generate this risk score for me? Why was my loan denied? Why did the system misclassify this image?*'.[18] In some situations, local explanations are essential to allow a user to act on an algorithmic output with confidence and justify his or her actions. For example, a bank officer will need to understand why a given transaction or customer has been classified as presenting a high risk for money-laundering before he or she sends a suspicious activity report to police authorities. Local explanations are also necessary to allow a person affected by a decision to challenge it, to prove that the output is incorrect, either because of incorrect input data, or because of a faulty model (one that is biased, for example). Local explainability presents the biggest challenge for providers and users of algorithms because the explanations often have to fit into a socio-technical system where humans must make decisions quickly.[19] To be meaningful, local explanations must focus on essential information, and take the cognitive limitations of the human user into account. Too much information will make explanations less useful.

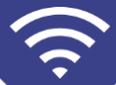

## 2.4 When Do Explanations Occur?

Local explanations may occur before a human decision is made, as in the case of an explanation given to a doctor before a medical diagnosis, or the explanation given to a police officer relating to an alert of potential terrorist activity requiring police intervention. This corresponds to a "human-in-the-loop" scenario. Local explanations may also occur after-the-fact, such as when a citizen wishes to contest an algorithmic decision. In some cases, local explanations may help the human supervising the system understand what the system is doing and intervene when necessary, such as when the system begins to shut-down, or when an aircraft navigation system begins to change its course. The human user may have a "stop" button that he or she is responsible for activating if the system begins to behave badly. This corresponds to a "human-on-the-loop" scenario.

To simplify, we can identify four main operational contexts for local explanations:[20]

- An explanation given to the data science teams during design and testing;

- An explanation given to the human user before he or she makes a decision based on the algorithmic output, a situation we will call *ex-ante* **"human in the loop"** explanations;

- An explanation given *after* a decision is made to permit a person affected by the decision, or an auditor, to challenge it, a situation we will call *ex-post* **"contestability" explanations**; and

- An explanation given to a human *simultaneously* while the algorithmic decisions are being made, to explain what is going on, a situation we will call **"human on the loop" explanations**.

The operational requirements for these four situations will differ. An explanation during design and testing is intended for expert data scientists, and time is not the principal constraint. An *ex-ante* human-in-the-loop explanation will need to help the human decision-maker choose a course of action (validate or not the output) in a relatively short time. An *ex-post* contestability explanation will be less constrained by time. The person wishing to contest the decision may have more time available to

---

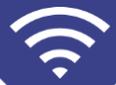

study the details of the explanation. A "human-on-the-loop" explanation needs to provide only critical information needed for the human to be satisfied that the system is functioning correctly.

Because local explanations are part of a dynamic process involving a human user or human subject who wants to understand and potentially act with regard to a particular algorithmic result, the problem of defining what constitutes a "meaningful" explanation is particularly challenging. Meaningfulness depends on the kind of audience and the operational context.

## 2.5 XAI Work Has Focused Principally on the Needs of Data Scientists

Explainability techniques have first arisen from a need for computer scientists to get a better grasp of the inner workings of the algorithms they were working on.[21] Typical tasks for them involve identifying performance issues and improving predictive accuracy. However, the techniques developed for these tasks are complex and necessitate a strong background in mathematics or computer science to be understood. On the other hand, systems designed for non-experts typically focus on a given work domain, for example, medicine, insurance or education. Users of such systems will need local explanations focused on answering specific tasks, such as identifying cancer cells in an MRI picture. Creating interactive systems that are usable, useful, and enjoyable to use is the focus of the HCI community, a subfield of research in computer science. Bertini and Lalanne 2009 showed that explainability techniques stemming from the AI community and the HCI community, and especially the information visualisation community, answered very different tasks, with different expectations about the user of their systems, and that both communities had a long way to go before bridging the gap between the two approaches.[22] Fourteen years later, this observation still holds true, even though some effort has been done with some attempts to bridge the two communities.

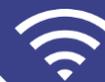

# 3. LEGAL REQUIREMENTS FOR XAI

A number of EU regulations and directives, whether already adopted or in discussion, impose different kinds of transparency obligations for algorithmic decisions. These transparency obligations are not presented as "explainability" *per se*, let alone as "global" or "local" XAI. But the transparency obligations contained in these texts are, in fact, forms of global or local XAI as we define them above, because they aim at rendering algorithmic processes and decisions understandable to humans. In addition to the XAI requirements flowing from EU regulations and directives, the CJEU imposed global and local explainability for algorithmic systems designed to detect terrorist risks, basing its decision on the EU Charter of Fundamental Rights.[23] XAI has also appeared in French laws dealing with algorithmic systems used by the State, which we will also review.[24] These requirements provide insight into why explanations are important to help achieve meaningful human control over AI systems, and/or to protect the individual rights of the persons, such as the right to contest algorithmic decisions. Our review of legal requirements will permit us to isolate five purposes of explainability: (i) contestability, (ii) empowerment and redressing information asymmetries, (iii) control over algorithmic performance, (iv) evaluation of individual algorithmic results, and (v) public administration transparency. These five purposes will be summarised in Section 3.8. The legal requirements also help us define four main categories of audiences for XAI: (i) data science team, (ii) human operator of the AI system, (iii) the citizen affected, and (iv) the regulator, judge or auditor. These audiences will be summarised in Section 3.9. Finally, in Section 3.10 we summarise the operational contexts already described in Section 2.4.

## 3.1. GDPR and Convention 108+

Both the GDPR and Council of Europe's Convention 108+ on the protection of personal data require data controllers to provide explainability for entirely automatic decisions.[25] The GDPR requires disclosure of meaningful information about the "logic involved" in the algorithmic decision,[26] while Convention 108+, which is not yet in force,[27] requires disclosure of the "reasoning underpinning" the decision.[28] There exists some academic debate as to whether these XAI requirements impose global or local explainability, or both.[29] However, the *purpose* of explainability in both these texts is to enhance **contestability** of decisions by the persons affected.[30] In our view, this supports the argument that both global and local explanations are required. This interpretation is supported by the official commentary to Convention 108+, which states that explanations should indicate how the model arrived at a particular decision regarding the affected individual.[31] In other words, Convention 108+ is

---

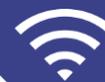

concerned with the particular algorithmic decision affecting the individual, and the explanation relating to that decision. Because the GDPR and Convention 108+, once Convention 108+ enters into force, will have equal normative weight and are supposed to be interpreted in a consistent manner, this points strongly in the direction of a local explainability requirement for both the GDPR and Convention 108+, which is an interpretation that also best serves the purpose of allowing individuals to contest individual algorithmic decisions.

In his March 16, 2023 opinion in the case *OQ v. Land Hesson SCHUFA Holding,*[32] the CJEU's Advocate General, Priit Pikamäe, said that the GDPRs provisions on explainability require "explanations sufficiently detailed on the method used to calculate the credit score and the reasons that led to a particular result".[33] Reference to a "particular result" clearly indicates that the GDPR requires, in the eyes of the Advocate General, "local" explanations. The Advocate General then continues: "In general, the data controller must furnish to the data subject global information, in particular on the factors taken into account for the decision process and on their respective importance at an aggregate level, which are also useful to contest any 'decision' in the context of Article 22(1) GDPR".[34] Here, the Advocate General refers to global explainability, although it is not entirely clear whether the "factors taken into account for the decision process and their respective importance at an aggregate level" refers to all credit scores generated by the system (global explainability), or the particular credit score at issue (local explainability). Reference to the "aggregate level" suggests that the Advocate General is referring to global explainability. The Advocate General's opinion confirms that the purpose of explanations is contestability.

The special provisions on explainability in the GDPR and Convention 108+ are limited to situations where the algorithm produces an algorithmic decision without significant human involvement. The Advocate General's opinion has broadened this to include credit scores that are then validated by humans. As we will see below in connection with recent CJEU decisions on anti-terrorism surveillance,[35] explainability, including local, may be required even in situations where there is significant human intervention in the final decision.

In addition to these specific XAI requirements for entirely automatic decisions, the GDPR imposes broad transparency obligations for any kind of processing of personal data, and these transparency obligations cover many of the same concerns as those addressed by XAI. In particular, the general transparency obligations imposed by Articles 13 and 14 of the GDPR cover much of the same information as would be included in global explanations imposed, for example, under Annex IV of the proposed European AI Act. In addition, it could be argued that the Convention 108+ and the GDPRs requirement of "fair" processing would require local and global explanations[36] and that such transparency is required to respect individuals' ability to choose freely how and when their data should be processed (**empowerment**),[37] another important purpose served by XAI.

---

[32] Opinion of the Advocate General of the CJEU P. Pikamäe dated March 16, 2023 in the Case C-634/21 OQ v. Land Hessen in the presence of SCHUFA Holding.
[33] Ibid., paragraph 58, translated from French by us.
[34] Ibid.
[35] Infra, Section 3.2.
[36] W. Maxwell, "Principles based regulation of personal data: the case of 'fair processing'", International Data Privacy Law, Volume 5, Issue 3, August 2015, Pages 205–216.

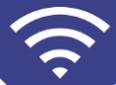

---

Summary of GDPR and Convention 108+ provisions:

- the main purpose of explanations is **contestability**, to allow individuals to effectively contest algorithmic decisions;
- a related, but less explicit, purpose is to preserve individual autonomy and **empowerment;**
- the GDPR and Convention 108+ require communication of the "**logic involved in**" or "**reasoning underpinning**" algorithmic decisions;
- the GDPR and Convention 108+ impose global, but also local, explainability, as pointed out by the CJEUs Advocate General in the case of *OQ v. Land Hessen/SCHUFA Holding*;
- these provisions apply only to decisions without meaningful human intervention, but CJEU case law (examined below) may extend explanations to cases where humans intervene in the decision process.

---

## 3.2. CJEU Decisions Applying the EU Charter of Fundamental Rights

Two decisions of the CJEU[38] have imposed global and local explainability in the context of algorithms used to detect risks of terrorism. Human operators, who receive and analyse individual alerts, must be able to understand the reasons for the alert, in particular to detect false alerts (false positives), and to detect possible discrimination. The CJEU emphasises that explainability is needed to permit effective **evaluation of individual results** by the persons reviewing alerts. A citizen who suffers the consequences of an alert must also be able to understand the reasons for the alert in order to exercise his or her right to effectively challenge the decision - **contestability**. According to the CJEU, machine learning algorithms are unable to provide an individualised understanding of the reasons for an alert, thus depriving the human operator of the ability to do his or her job of detecting errors correctly, and depriving the citizen from his or her right to challenge the action in court. Therefore, the Court has said that algorithms for these applications should be based on "**pre-determined models and criteria**",[39] and that machine learning algorithms should be prohibited for anti-terror detection systems.[40] The CJEUs decisions mean that local explainability, of the *ex ante* human-in-the-loop kind, is a fundamental right protected by the Charter whenever the decision leads to serious consequences for the rights and freedoms of individuals. Local explainability, of the *ex post* contestability kind, is a fundamental right protected by the Charter, whenever the individual's right to an effective remedy would be compromised without it. The requirement of global explanations is reflected in the Court's requirement that the system be built around "pre-determined models and criteria". According to the

---

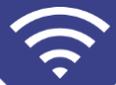

Court, these criteria must be shown to be effective in identifying real terrorist threats. This aspect of explainability aims at ensuring **control over system performance**.

---

Summary of CJEU decisions based on the EU Charter of Fundamental Rights:

- in a context with high risks for fundamental rights such as the detection of threats of terrorism, local explainability is necessary to permit the human operator in charge of human control to **evaluate individual results**, to detect errors and potential discrimination in alerts;
- local explainability is also necessary to ensure **contestability** of individual decisions by the persons targeted;
- systems should be based on "**pre-determined models and criteria**", which guarantees global explainability and the ability to **control system performance**.

---

## 3.3. The Proposed European AI Act

The European AI Act, in the version proposed by the Commission,[41] would impose extensive transparency requirements on high-risk algorithmic systems, but these requirements relate to **global** explanations, through technical documentation and user instructions. Local explanations are not mentioned in the text, at least not explicitly. Local explainability may be indirectly required, however, through Article 14-4(c) of the draft regulation, which requires that systems be designed so that the human(s) in charge of oversight are "*able to correctly interpret the high-risk AI system's output, taking into account in particular the characteristics of the system and the interpretation tools and methods available*". Understanding a particular algorithmic output will generally require local, not only global, explanations. When preparing a risk assessment and proposing mitigation measures, providers of high-risk AI applications will likely find that some form of local explanation is required, not only to guarantee effective human oversight, but also to permit contestability. However, compared to global explanations, which are specified in great detail in Annex IV of the draft regulation, local explanations are dealt with only indirectly, hence our suggestion in Section 6 to provide more guidance to providers and users of high-risk systems in how to design meaningful local explainability mechanisms. The proposed AI Act imposes global explanations in order to permit **control over system performance**, to discipline AI providers to disclose the weaknesses and use limitations of the system, expose characteristics of the system to outside scrutiny, and permit users to apply the systems safely. The purpose of local explanations, which again are not directly referenced in the proposed AI Act, is to enable effective human oversight over algorithmic outputs, for example, **evaluation of individual results**.[42] Contestability is not expressly mentioned in the AI Act draft.

---

The proposed European AI Act:

- emphasises global explanations, with the objective to permit effective **control over system performance** as well as **evaluation of individual results**;
- local explanations of individual results are not expressly addressed in the AI Act, but will likely be required to permit humans in charge of oversight to "**correctly interpret**" algorithmic output.

## 3.4. Platform Regulation

As we will see in this section, several EU laws protecting consumer and business users of platforms require explanations regarding the "main parameters" of ranking algorithms and recommender systems. The most detailed provisions on XAI are contained in the **Platform to Business ('P2B') Regulation**,[43] which has been supplemented by Guidelines[44] published in 2020 by the European Commission. The P2B Regulation and associated Guidelines adopt a user-focused approach to XAI insofar as they require that the information on algorithmic parameters used in ranking permit the average business users of the platform to have an "**adequate understanding**" of the ranking system.[45] This understanding should permit the users of the platform to improve their own positioning in the system and enhance predictability. The P2B Regulation and Guidelines also require some justification for the **choice of the main parameters and their relative importance**, suggesting that the main parameters, and their weighting, must follow a coherent thought process rather than a machine-created model.[46] The Guidelines underline the need to avoid excessive information, which would destroy the intelligibility of the explanation for the user.[47] Also, the information on the main ranking parameters should not allow bad faith manipulation of the service.[48]

The transparency obligations in the P2B Regulation and Guidelines aim at inserting rationality into ranking systems so that users can exercise some control over their own fate, by taking steps to improve the ranking of their service with regard to certain important ranking criteria. These provisions aim at enhancing individual **empowerment**, the importance of which has been highlighted in the literature on counterfactual explanations.[49] Counterfactual explanations generally focus on the smallest possible changes within the control of the user that would lead to a different classification

---

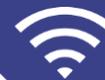

by the system, such as "if your salary were increased by x or more, your credit score would improve to the low risk category". One role of explanations is to permit individuals to adjust their behaviour, which is an important element of individual autonomy and empowerment.[50] Counterfactual explanations permit this.

The revised **Consumer Protection Directive** also contains provisions requiring disclosure of the "main parameters" of ranking, but this time aiming at protecting consumer, as opposed to business, users of ranking services.[51] The operator of the platform must provide to the consumer "**the main parameters determining the ranking of products presented to the consumer as a result of the search query and the relative importance of those parameters, as opposed to other parameters**".[52] This language closely tracks that in the P2B Regulation. One can imagine, therefore, that a number of elements contained in the P2B Regulation Guidelines would be transposable to the Consumer Protection Directive, such as the need to avoid too much information so that information remains meaningful. However, the "empowerment" rationale is less present in the context of consumer searches than in the context of business users' offers. A consumer should understand why certain offers are prioritised in order not to be misled by the ranking. But unlike a business user, a consumer does not generally need to understand ranking systems in order to improve his or her own positioning in the system. The protection of the consumer aims above all at **redressing information asymmetries**.

The **DSA** requires disclosure of the **main parameters** of recommender systems and of systems that present personalised advertisements.**[53]** The DSA requires not only disclosure of the "main parameters" for recommender systems, in clear and intelligible language, but also **the criteria that are most significant and the reasons for the relative importance** of those parameters.[54] The Commission will encourage the development of voluntary standards on "choice interfaces and **presentation of information on the main parameters of different types of recommender systems**".[55] Like the P2B Regulation, the DSA requires disclosure of the "reasons" for the relative importance of the main parameters, suggesting that, like the P2B Regulation, the choice of parameters and their weighting must follow some logical human thought process, rather than a purely machine-derived function based on observed statistical correlations that may or may not correspond to a valid reason. For example, one of the parameters selected by an algorithm to recommend certain content may be the type of internet browser used, or the speed of typing. Yet these two features may not lend themselves to giving a valid "reason", if what is meant by "reason" is a logical and defendable justification for a given choice of parameters. It will be left to regulators, and ultimately the CJEU, to determine if the obligation to provide "reasons" implies a requirement that the reasons have some logical relationship to the outcome that the system is designed to predict.[56]

---

[50] Information Commissioner's Office (ICO) and The Alan Turing Institute, Explaining Decisions Made With AI, Guidance Document (2022).
[51] Directive (EU) 2019/2161 of the European Parliament and of the Council of 27 November 2019 amending Council Directive 93/13/EEC and Directives 98/6/EC, 2005/29/EC and 2011/83/EU of the European Parliament and of the Council as regards the better enforcement and modernisation of Union consumer protection rules.
[52] New article 7-4(a) of Directive 2005/29/EC, created by Directive (EU) 2019/2161 of the European Parliament and of the Council of 27 November 2019 amending Council Directive 93/13/EEC and Directives 98/6/EC, 2005/29/EC and 2011/83/EU of the European Parliament and of the Council as regards the better enforcement and modernisation of Union consumer protection rules.
[53] Regulation (EI) 2022/2065 of 19 October 2022 on a Single Market for Digital Services and amending Directive 2000/31/EC ('DSA').
[54] DSA, Article 27.
[55] DSA, Article 44-1-i.
[56] On the distinction between explanations and reasons, see C. HENIN and D. LE MÉTAYER, "A Framework to Contest and Justify Algorithmic Decisions," 2021 *AI and Ethics*, Springer, 1, pp. 463-476.

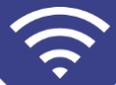

The rationale of the provisions in the DSA on recommender systems and advertising is both to avoid misleading individual users, and to **empower** individuals to control their own information environment by exercising influence over recommender systems.[57] The provisions of the DSA relating to content moderation systems will be discussed in the following section.

---

The P2B Regulation, the Revised Consumer Protection Directive and the Digital Services Act:

- require global explanation of the **main parameters, their relative importance, and the reasons for their choice;**
- the purpose is to promote **empowerment** of users, for example, the ability to take action to affect their own algorithmic fate;
- the purpose of the consumer protection directive is also to **redress information asymmetries**.

---

## 3.5. Regulation of Automatic Content Moderation Systems

The Online Terrorist Content Regulation[58] and the DSA[59] impose varying forms of explanations for automatic content detection and removal tools. The Online Terrorist Content Regulation imposes both a "**meaningful explanation of the functioning**" of the algorithm as a whole (global explanation), and a local explanation for a particular removal.[60] For example, the regulation requires that the platform operator give the "**reasons for removal**" of a given piece of content, the reason for an individual removal being a form of local explanation.[61] The DSA, already discussed above in the context of recommender systems, requires a "qualitative description, a specification of the precise purposes, indicators of the accuracy and the possible rate of error of the automated means used in fulfilling those purposes, and any safeguards applied", which is a form of global explanation.[62] In addition, the platform provider must give **reasons for a specific content removal action** (local explanations) in order to permit the user to challenge the decision.[63] The aim of these provisions of the DSA is both **empowerment** - to permit users to understand in advance what content is likely to be filtered and therefore avoid uploading such content - and **contestability** - to permit users to understand the reasons for a content removal decision in order to effectively contest the removal. The DSAs requirement to describe the precise purposes, indicators of accuracy, and possible rates of error may require extensive documentation by the platform provider, which may be difficult to understand for the user. Moreover, the purposes and rates of error of the system could change over time as content moderation systems, and their error rates, are adjusted to adapt to new circumstances. These issues may have to be dealt with by the European Centre for Algorithmic

---

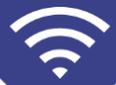

Transparency, an agency of the European Commission dedicated to studying algorithmic transparency issues.[64]

---

The Online Terrorist Content Regulation and the DSA:

- impose **both global explanations** – "explanation of the functioning", "general description" – and **local explanations**, for example, the reasons for a particular algorithmic removal;
- the DSA imposes a description of the precise purposes, indicators of the accuracy, and the possible rate of error of the algorithm;
- the purposes are to enhance user **empowerment** to understand and adapt to the platforms' rules, and **contestability** to challenge individual removal decisions.

---

## 3.6. The Proposed Platform Workers' Directive

The proposed directive to protect digital platform workers requires disclosure of the **main parameters** of algorithmic systems and their **relative importance**.[65] Workers must also receive information on how their personal data and behaviour influence automatic decision systems. This form of explanation is "global" in the sense that it describes the system as a whole, and its objective is principally **empowerment**, to allow platform workers to have some understanding of the system in order to exercise some control over their own destiny. In addition, the proposed directive requires written **reasons** for individual decisions in order to permit the affected worker "**to discuss and to clarify the facts, circumstances and reasons having led to the decision**" with a human representative of the platform.[66] This form of explanation is "local" because it focuses on an individual decision, and its objective is **contestability**, to ensure workers have an effective remedy with regard to decisions affected by algorithmic output. The proposed directive also focuses on human monitoring of automated systems by qualified staff, a measure that will require, to be effective, both global and local explainability.[67] The purpose of the provision of human monitoring is to ensure **control over system performance**. Finally, Article 9 of the draft directive focuses on information and consultation with workers' representatives regarding automatic monitoring tools, an article that will also require global explainability on the objectives and functions of the algorithm.

---

The proposed Platform Workers' Directive:

- requires disclosure of the **main parameters and their relative importance**, for example global explainability;
- local explainability is required insofar as the platform must give **reasons** for individual decisions;
- global and local explainability will be required to permit effective human monitoring of the system;

---

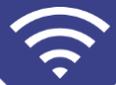

> - the information and consultation of workers' representatives is necessary, which will require global explainability;
> - the main purposes of explainability are **empowerment** of platform workers, **contestability, and control over algorithmic performance** by persons in charge of human monitoring.

## 3.7. French Code of Relations Between the Public and the Administration

For any individual decision based on an algorithmic processing operation, the French Code of Relations between the Public and the Administration requires disclosure, at an individual's request, of the **rules defining the processing** as well as the principal characteristics of its implementation.[68] More particularly, individuals have the right to obtain information relating to: (1) the degree and form of contribution of the algorithmic system to the decision; (2) the data processed and their sources; (3) **the processing parameters, and, as the case may be, their weights, as applied to the situation of the individual**; and (4) the operations performed by the processing.[69] The French code requires disclosure of <u>all</u> parameters and their weights in an individual decision, not just the main ones. These enhanced "local" explanations for public decision-making that rely in whole or in part on algorithmic systems are necessary both for **contestability** and for **public administration transparency**, protected by Article 41 of the Charter[70] and by corresponding provisions of national constitutions. These provisions apply to any individual decision based, even in part, on algorithmic results, even if the decision is made by a human. This contrasts with the GDPRs special provisions on explainability, which apply only to decisions without significant human involvement.

> The French Code of Relations between the Public and the Administration:
>
> - requires both **global and local** explanations;
> - requires disclosure of the **processing parameters, and, as the case may be, their weights, as applied to the situation of the individual**;
> - the purpose is to enhance **contestability**, as well as **public administration transparency**, a separate right protected by the EU Charter.

## 3.8. The Five Purposes of XAI Resulting From Legal Requirements

The review of legal requirements permits us to identify five **purposes of explainability**:

**Contestability**: Explanations, global and local, are required to allow individuals to effectively challenge decisions, whether made by public or private actors. Contestability of decisions made by public

---

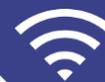

administrations is a fundamental right protected by the Charter. Contestability of decisions made by private actors is protected by several legislative instruments mentioned above, including the GDPR.

**Empowerment and redressing information asymmetries**: Global explanations are required to allow users, whether businesses or individuals, to understand, and have some choice and influence over how they interact with the system. For individuals, autonomy is a fundamental right protected by the Charter, the GDPR and Convention 108+. For businesses, empowerment is part of fair business practices, designed to level the playing field in situations of economic or informational imbalances. Redressing information asymmetries also avoids unfair commercial practices, and is an element of consumer protection law.

**Control over system performance**: Global explanations are necessary to discipline providers of AI systems to test and disclose all aspects of the system that might impact its safety and quality in operation. Global explanations also help users and persons in charge of human oversight to know the limits of the system and thereby avoid accidents.

**Evaluation of individual algorithmic results**: Local explanations are necessary to help humans effectively oversee and control individual algorithmic output. When bad output can affect individual rights, the need to ensure the quality of individual decisions becomes a fundamental right protected by the Charter.

**Public administration transparency**: Global and local explanations are necessary to ensure the public's right to transparency and good administration, protected by the Charter and national constitutions.

Other objectives of XAI exist, such as permitting data scientists to improve their models, or permitting scientists to improve scientific discovery.[71] But these objectives are generally satisfied by forces of the market and are therefore not the target of regulation.

## 3.9. The Four Audiences of XAI

The review of legal requirements also permits us to identify four principal audiences for explanations:

**Data science teams**: as mentioned above,[72] XAI has been designed foremost to help data scientists understand models in order to improve them. This category includes all the persons who want to understand an algorithm in order to test and improve performance, identify and reduce biases, and/or to make the system more transparent.

**Human operator of the system**: one of the main intended recipients of explanations is the human in charge of overseeing the system, either in the capacity of "human-on-the-loop" in charge of monitoring the system and stopping it if necessary, or in the capacity of "human-in-the-loop" in charge of validating every algorithmic output before action is taken.

---

[71] See infra, section 4.4.
[72] Supra, Section 2.5.

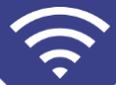

**Persons affected by the algorithmic decision**: this category includes both businesses and individuals who are affected by an algorithmic decision. It also includes organisations that represent the collective interests of individuals affected by algorithms, such as workers' or consumers' organisations.

**Regulator, judge or auditor:** this category includes any person who is in charge of evaluating, usually after the fact, the performance of the system or the appropriateness and legality of a particular output. This category could also include independent researchers looking for bias. It may also include persons within the system operator's company in charge of dealing with complaints from the persons affected by algorithmic decisions.

## 3.10. The Four Operational Contexts

Finally, as noted above,[73] there are four main operational contexts in which explanations are needed. This categorisation is an oversimplification, but it helps identify the different explanation constraints that may be present.

**Design and testing of the system**: as noted above, XAI is an important tool for the teams designing and testing algorithms. These teams will be experts and will not be subject to short time constraints.

**Human-on-the-loop**: this situation corresponds to the human operator overseeing the system to make sure it is operating properly. A dashboard would typically be used to alert the human supervisor of possible anomalies, and the human supervisor will generally have a very short time to understand what is going on, and eventually take remedial steps such as activating a stop button. This situation is comparable to a human pilot receiving an alert from the automatic pilot system that something is wrong.

**Human-in-the-loop**: this situation corresponds to a human evaluating each and every algorithmic output to verify its relevance before taking action, for example in case of a facial recognition match, or in case of an alert of possible money-laundering. The human will have very little time to make his or her evaluation.

**Ex-post challenge, investigation**: in this situation, the person affected, a regulator, an auditor, or a data scientist will want to understand what happened, in order to challenge the decision, or improve the system. There will be fewer time constraints in this situation.

## 3.11. Summary Table of Legal Requirements

The table below summarises the legal requirements described in this Section 3. Some of these texts may apply simultaneously to the same digital activities, highlighting the difficult task of companies trying to interpret these provisions and adopt compliant explainability approaches.

---

[73] Supra, Section 2.4.

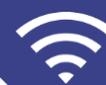

| Source of law | Purposes | Key terms |
|---|---|---|
| CJEU (La Ligue des droits humains) | Contestability, control over system performance | "pre-determined criteria"; "able to understand how those criteria and those programs work" |
| P2B regulation; Consumer Protection Directive | Empowerment and redressing information asymmetries | "main parameters", "reasons for the relative importance", "adequate understanding", "meaningful explanations must take into account nature, technical ability, and needs of average users" |
| Data protection (GDPR, Conv. 108+) | Contestability, empowerment | "meaningful information about the logic involved", "knowledge of the reasoning underlying" |
| Platform Workers Directive | Contestability, empowerment | "main parameters and relative importance", "grounds for decision" |
| Content moderation/recommendation on platforms (DSA, Regulation on Online Terrorist Content) | Contestability, empowerment | "main parameters", "meaningful explanations of the logic", "meaningful information", "reasons". DSA: "precise purposes", "indicators of the accuracy and the possible rate of error" |
| Proposed AI Act | Control over system performance | "fully understand the capacities and limitations", "correctly interpret output" |
| French administrative code | Public admin. transparency | the "parameters and weights" |

Table 1: Summary of Legal Texts Requiring Explainability of Algorithms

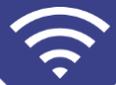

# 4. USER-CENTERED METHODOLOGIES FOR EXPLAINABILITY

## 4.1 The XAI Community Has Not Yet Fully Embraced User-Centred Design Methodologies

The foregoing sections have given us some understanding of the purposes, the intended audiences and the operational contexts in which explanations might be required. Once we know the purpose, the audience, and the operational context of an explanation, the only way to define "meaningful" XAI is to study how explanations work on people. Phillips et al. define "meaningful" explanations as ones that are understandable to the intended recipient.[74] Explanations that are understandable by their intended recipients are "meaningful", those that are not understandable are not "meaningful". Studying the effects of explanations on people is the subject of HCI studies, which focus on the design of computer technology and, in particular, the interaction between humans (the users) and digital systems. Practitioners in HCI typically take a user-centred point of view when designing digital systems and their user interfaces. The HCI research community was thus directly interested when explainability issues of AI-based systems appeared.

Typically, HCI practitioners rely on so-called user-centred design methodologies when designing interfaces.[75] These methodologies work in iterative, cyclic steps, with 4 to 6 different stages depending on the methodology. These stages usually involve first setting the stage by analysing who the audience is, what the expected tasks of the future users are, what (if any) regulations need to be followed, and more broadly exploring the expected context of use. Based on this initial analysis, the design process usually starts with a number of ideation cycles, which will generate initial prototypes. These prototypes are then iteratively refined and reviewed, either by the design team, by experts in user experience or by end users. User-centred design approaches succeed because they place the end user at the focal point. This guarantees that the final product responds to the needs and wishes of the target audience. Most major companies in the IT field have adopted variants of user-centred design methodologies for the design of the interfaces of their products.

In contrast, as mentioned above, the field of explainability of AI-based systems has not yet fully embraced this approach. This is due to a number of different reasons, including the novelty of the field or the fact that most research still done currently revolves around the algorithmic aspect of explainability. Many explainability techniques have been developed by AI practitioners and engineers, so that they can have a better vision of how their algorithms perform.[76] However, the resulting explainability techniques are ill-suited to be deployed in systems intended for a general audience. As an example, one may consider t-SNE, a visualisation technique developed to represent high-dimensional data, such as the models created by machine learning algorithms.[77] An example of a visualisation created with t-SNE can be seen in Figure 1. The visualisations created by t-SNE are already considered difficult to interpret for an expert in machine learning, and a user without a university degree in computer science or mathematics will have little chance to comprehend the representation. Such explanations cannot be considered "meaningful" in light of the purposes and audiences mentioned in Section 3.

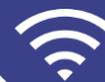

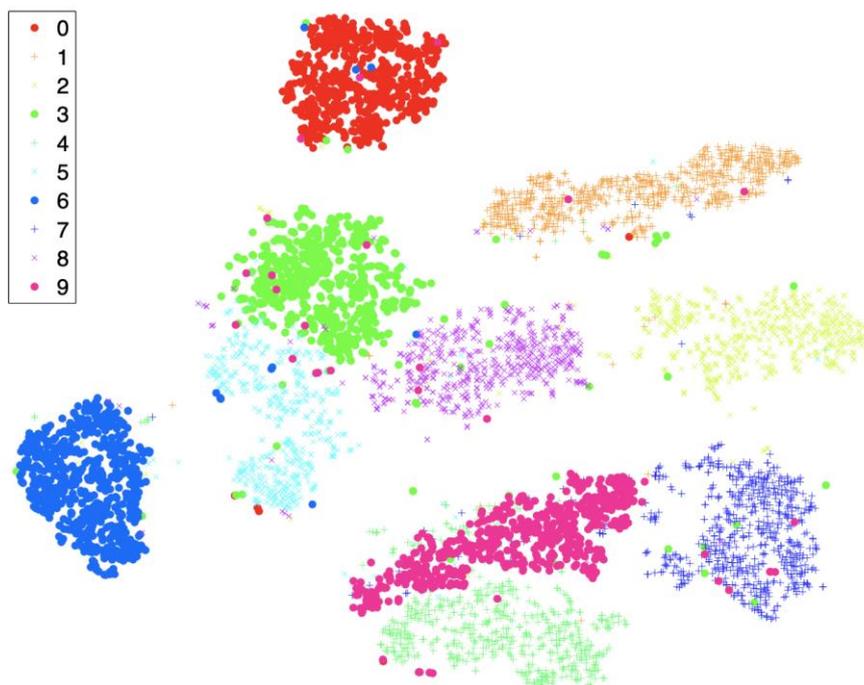

Figure 1: An example visualisation using the t-SNE technique[78]

When taking a user-centred design vision on explainability techniques, the main question to ask becomes the following: "how can the end user of an AI-based system get enough understanding of the system so that she may use the system to resolve her tasks confidently?". This initial question opens a number of sub-questions, for example:

- Who is the end user of the system?
- What are her tasks?
- What, and how much does she need to know to use the system with confidence?
- What is the best shape this information may take to be understood efficiently, based on the user's profile?

In practice, these questions help designing an AI-based system which makes sense to its end users, brings acceptance and trust to the system. Explanations are not, however, a panacea. In some cases, they may defeat their intended purpose.

---

[78] Supra footnote 76.

## 4.2 Some Explanations Can Make Users Complacent

Some explanations make users worse off, by increasing their complacency and reducing their vigilance compared to users who receive no explanations. Brandimarte et al. showed that giving increased control to users over their personal data generates excessive confidence and yields worse outcomes for individuals than for the group who had little control over their personal data.[79] More recently, HCI researchers have shown that explanations can increase the likelihood of humans wrongly following algorithmic recommendations.[80] Explanations can exacerbate human cognitive biases, making users worse off. Some explanation techniques can try to counter those biases. For example, some explanations may seek to point out reasons an algorithmic recommendation should *not* be relied on, in order to heighten users' vigilance and reduce automation bias.[81] In one study involving explanations of investment proposals, the group that received the explanation made more bad decisions than the group that received the explanations, which tends to confirm that explanations can make users over-trusting.[82] These results underline that explanations should never be considered as an objective in themselves, but as a means to achieve other objectives. The value of an explanation must be measured by its level of achievement of those other objectives.

Below are two illustrations of explanations in context.

## 4.3 Use Case 1: Education

In Abdi et al. 2020, the authors present an educational recommender system.[83] Such systems help students learn a given topic by adapting the resources and activities to best suit the student's profile. However, in many cases, these educational recommender systems work like black boxes, without giving the students any rationale about the recommendations that are given to them. The authors explored an open learner model strategy, where the model representing a learner is externalised in a way that lets the learner get a better understanding of the reasons the model recommends a given content. The authors used the Elo rating system as a way to quantify the knowledge of a learner in a given field. An illustration of the system created by the authors is illustrated in Figure Y. As can be seen, a set of recommendations of new lessons to undertake by the learner are suggested below a graph visualisation showing how well the learner performed in several knowledge units. The colours of the bars help the learner quickly spot the knowledge units where more work is required, and how well she compares to the rest of her class. Finally, the user can select a given knowledge unit to focus on the relevant topics that may be studied to enhance her knowledge.

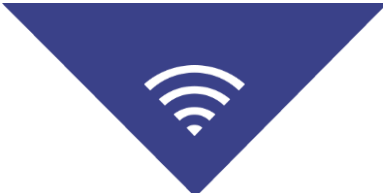

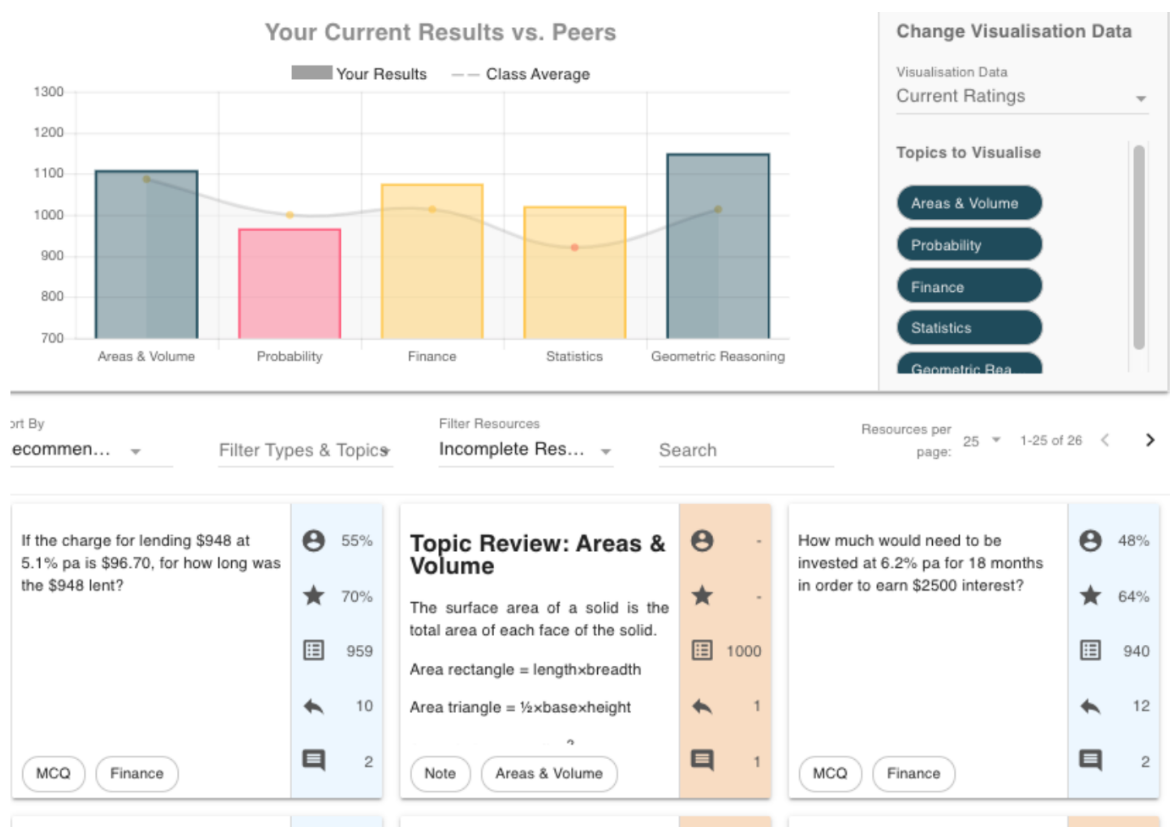

Figure 2: A Screenshot of the Educational Recommender System Presented in Abdi et al. 2020[84]

In the case of this system, if we consider the four questions enumerated above, the users are clearly identified: learners using an educational recommender system. The tasks are also well defined: the learner is looking for topics that can help her get better in a number of knowledge units. The visualisation that was selected helps the user pinpoint knowledge units to focus on without drowning her in needless information. The performance of the rest of the class is also a helpful indicator, yet still protects the private data of the other students. Finally, the graph visualisation is pretty straightforward to understand and helps pick the most relevant topics to study, letting the learner make a better, informed decision.

---

[84] Supra footnote 81.

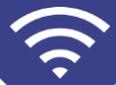

## 4.4 Use Case 2: Explainable AI for Scientific Discoveries

This use case is a bit different than the previous one, in that it is less about usability and making sure that the user can get an easy understanding of some aspects of an AI-based system, and more about letting an expert user dive into specific information of a model. However, as will be seen, the four questions presented above are still relevant.

Scientists were among the first users of AI-based systems, and particularly machine learning-based systems. As Roscher et al. 2020 mention, "In the natural sciences, the main goals for utilising ML are scientific understanding, inferring causal relationships from observational data, or even achieving new scientific insights".[85] Machine learning techniques are especially interesting as they can interpret large quantities of data quickly and efficiently. As an example, astrophysicists have been using machine learning to detect potential exoplanets candidates in datasets captured by telescopes.[86] For such use cases, explainability can have a key role to play in helping scientists coming up with scientific outcomes. These can stem from explanations of output results of a given machine learning algorithm, by predicting scientific parameters and properties, or by explaining models of the algorithms. In these cases, tools used will focus much more on the exploration of data and models, requiring at least some expertise in how machine learning techniques work. A visualisation such as t-SNE may make sense in such a use case.

These different use cases show the diversity of cases where explainability techniques for AI-based systems would be useful. They also show that different explainability techniques can be applied to different audiences, for different tasks, and that different representations will be better suited depending on the task and the context. Tailoring the explanations to the right audience and the right tasks is a matter of choices and trade-offs. As of now, no "one-size fits all" technique exists, and it is doubtful any will appear in the future. As an illustration, the HCI community has come recently to the realisation that, in some cases, purely text-based explanations are better than the visual ones illustrating the different use cases above.

In some way, explainability techniques for AI-based systems is a domain still in its infancy. When graphical user interfaces were conceived and popularised at the beginning of the 1980s, it took a dozen years for HCI scientists and practitioners to find and describe methodologies and best practices to create usable interfaces, and then another 10 years before the impact of emotional aspects was realised, paving the way for the appearance of user experience-oriented practices. As of now, explainability techniques are still tailored in a per-case situation, and it might be still a few years before some more general guidelines and principles appear, helping in the design of explainable AI-based systems. What may already be said is that, probably, variants of user-centred methodologies could help in the first stages of the design of such systems. Furthermore, these methodologies lend themselves well to the integration of regulation-based guidelines.

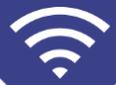

## 4.5 Defining a "Meaningful" Explanation

As noted above,[87] the EU General Data Protection Regulation ('GDPR') gives data subjects the right to "meaningful information about the logic involved" behind algorithms for entirely-automated decisions. However, what "meaningful" is supposed to entail was not described further in the text, leaving its interpretation open. "Meaning" is closely related to language, and is described as "the thing that is conveyed",[88] "something meant or intended",[89] or a "significant quality".[90] "Meaningful" is itself described as "having a meaning or purpose",[91] or "significant".[92] The French language version of the GDPR uses the term "*information utile*", showing the link between meaningfulness and the utility of information for the recipient. It is already interesting to note that "meaningful information" relates to the language or, at least, to the communication modality, to information having some significance or purpose, and even to intention. Finally, meaningful information may be information having a "significant quality". As one may notice, the spectrum of what "meaningful information" may cover is relatively large. Even worse, "meaningful information" is not a consistent quality when referring to information presented on a computer interface, as would most likely be the case with "meaningful information about the logic involved" behind a given algorithm. Meaningfulness is closely linked to the effect on the recipient: did the recipient of the explanation understand it?[93]

In the field of computer-human interaction, multiple works,[94] have explored the concepts of meaning, and what meaningful interfaces may have as specific qualities. In the process, researchers relatively quickly realised that what is considered meaningful may vary from one person to the other. As an example, the level of education, job expertise or digital literacy are all aspects which will influence how some information is perceived. As an example, one may consider Figure 1 above. The t-SNE visualisation technique has been considered for many years the gold standard of explainability techniques for machine learning. Each point of Figure 1 corresponds to a data point characterised by a number of dimensions. These dimensions represent features of the data sample. As an example, a country could be characterised by the number of inhabitants, the land size, the GDP, and so on. The number of dimensions can easily rise into the hundreds, thousands or even much more, depending on the dataset. As hundreds of dimensions cannot be easily reduced to the two X/Y dimensions of a screen, visualisation techniques like t-SNE try to cluster together data points for which their dimensions are comparable. As such, these techniques are definitely designed to convey meaning, however without a strong understanding of how machine learning typically works, linear algebra, N-dimension spaces as well as the specific dataset used, it can be next to impossible to gather any meaning from a t-SNE visual representation. As such, t-SNE is a good example of a tool typically used by engineers and computer scientists when they develop and train a new machine learning-based system.

---

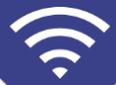

In comparison, Figure 2 presents a system targeted at students. The designers of the system made sure that the information is not overwhelming, and that a passing knowledge of the education system should be enough to understand the most relevant information presented. One might ask though whether the information is sufficiently detailed to be really "meaningful". A professor might want a more detailed view of the precise topics where a student failed to help him improve, for example.

These two examples should help illustrate that meaningfulness is relative to a number of different aspects. Beyond the role, knowledge, job, expertise and overall profile of a user, it has been observed that the context of use (where, when, how) and the task of the user had also a potential impact. Finally, even the current mood of the user might have subtle influences on how "meaningful" specific information may be judged.[95] "Meaningful" is thus relative, contextual, and may vary strongly from one user to the other, from one task to the other, and from one context to the other.

## 4.6 Measuring "Meaningfulness"

Current work in XAI focuses on metrics for measuring an explanation's quality.[96] We focus below on only one aspect, which is an explanation's "meaningfulness". As described in the previous section, meaningfulness is not definite or invariable. On the contrary, some information may be more or less meaningful depending on the user, on the context, on the task. Assessing how meaningful some information or explanation is needs to take into account these variables. If meaningful information is to be assessed, some kind of measures are needed. This is where the field of human-machine interaction may help. As described before, practitioners in human-machine interaction have been developing methods to assess various aspects of usability and user experience of digital interfaces, such as efficiency, ease to understand or aesthetics. Meaningfulness has gathered interest more recently in the community, with the growth of AI-based techniques deployed in the field.

The question of how to measure meaningfulness was raised in a 2019 workshop at one of the top conferences of the human-machine interaction domain.[97] Dimensions of meaningfulness were described according to three different models. One of these models, taken from Martela & Steger (2016),[98] is of particular interest, as it describes three main levels leading to meaningful interactions. In what follows, the word "product" is used to describe systems presenting information or explanations. The first level, "coherence", involves clarity and sense-making, with questions such as:

- Is it clear to the user how the product connects to their life?
- Does the product make sense to the user?
- Does interacting with the product feel right?

The second level, called "purpose", describes how the user experience brings motivation to the user. Questions revolving around purpose can be:

- Does the product help users identify personally important goals?

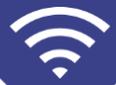

- Does the product help users set manageable smaller goals to reach personally important goals?
- Does the product support users in reaching and achieving those goals?

Finally, the third level tackles "significance":

- Does the product matter to users beyond the momentary interaction? How so?

In the focus of the present document, the first level, "coherence", is of particular interest, as is the level "purpose". The third level, "significance", is of less interest as it relates to issues more linked to general user experience. However, it should not completely be set aside, as interacting with any digital system will have multiple diffuse impacts on users.

Meaningfulness in the context of XAI can thus be resumed to the following subquestions:

- Does the system make sense to the user?
- Does interacting with the system feel right?
- Does the system support users in reaching and achieving pre-set goals, such as the goal of permitting contestability?

Hsiao et al. (2021)[99] describe in more detail seven cognitive metrics linked to these questions, of particular interest when measuring and assessing AI-based explainable systems:

- Explanation Goodness
- User Satisfaction
- User Curiosity/Attention Engagement
- User Trust/Reliance
- User Understanding
- User Performance/Productivity
- System Controllability/Interactivity

All these cognitive metrics can be measured in one way or another. Qualitative as well as quantitative techniques, frequently borrowed from fields of social sciences such as psychology or ethnography, have been adapted and have shown their strength in giving interesting results when assessing measures such as the one presented above. However, these techniques all have one aspect in common: they all rely on users as the final judges.

A typical evaluation session involving users will start with an already existing version of an XAI system, such as the ones illustrated in Figures 1 and 2. A number of users will then be confronted with the system, manipulate it, and try to reach pre-set goals. After the session, the users will then fill out one or multiple questionnaires and be interviewed on how they experienced the overall session. This is obviously a rough outline of a generic usability evaluation session, and playing with the fidelity of the interface, the number of users, the types of questionnaires used, the topics touched upon during the interview or the types of tasks will allow the evaluators to focus on one or multiple of the cognitive

---

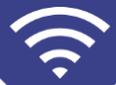

metrics described above. However, these techniques provide answers to the question of assessing meaningfulness of XAI systems. Detailing further these evaluation techniques is out of scope of this report, but Hsiao et al. (2021)[100] bring interesting advice and pointers for the reader wishing to delve deeper.

One last question remains: how effective are these usability evaluation techniques at assessing precise, specific measures of meaningfulness? The honest answer would be: "varying". One has to remember what we mentioned in the former section: meaningfulness is thus relative, contextual, and may vary strongly from one user to the other, from one task to the other, from one context to the other. This has relatively strong implications:

● One cannot expect complete reproducibility of the results of an evaluation session, as the overall user experience can and will change from one user to the other.
● Evaluation results will be of a statistical nature, with more or less large error margins.
● Evaluation results will be as good as how well the evaluation session was planned and run, and planning and running well an evaluation session asks for some expertise in user-centred design.

These limitations being stated, we also need to mention that evaluation sessions as described here give surprisingly good and even accurate results, when well-planned and well-run. In some cases, only a handful of users (4-5) can already be enough to assess reasonably well how meaningful a system is. In some other cases, more extended approaches with dozens or even hundreds of users might be necessary to reach high enough statistical accuracy.

In general, when trying to assess whether an XAI system is "meaningful" for its users, a "best efforts" approach should be privileged, relying on user-centric evaluation techniques, with well-defined tasks related to the systems, users with expertise linked to the system being evaluated, and honest analysis of the results. This best efforts approach seeks meaningfulness for an average user, which means exploring techniques that work best on most but not all people. This best efforts approach contrasts with the typical approach of regulators, which is to put themselves into the shoes of an individual interacting with the system and judge for themselves whether the information is sufficiently meaningful for them. The opinion of the regulator, putting herself into the shoes of a user, may not coincide with the conclusions of testing using user-centric design approaches. We suggest therefore that regulators, when they assess whether an explanation is meaningful, take due account of user-centric design methodologies undertaken by the provider or user of the algorithm.

---

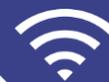

# 5. COMBINING THE "LEGAL" AND "USER-CENTRIC" APPROACHES INTO A COHERENT METHODOLOGY FOR XAI

As we have seen, the problem of creating meaningful explanations for algorithmic decisions depends on the purpose of the explanation, the recipients of the explanation, and the operational context in which the explanation is given. The user-centric approach used in HCI has ways to test various criteria of "meaningfulness" of explanations, but these approaches will show differences in "meaningfulness" depending on the characteristics of the users. There will be no solution that is meaningful to everyone, which puts enterprises in a difficult situation with regard to their regulatory obligation to provide "meaningful" explanations.

Some form of regulatory guidance and harmonisation on meaningful explanations is needed. The focus of regulatory guidance should be on the methodology to be used, not the precise XAI technique that should be applied. A harmonised methodology for designing "meaningful" AI explanations is similar to the harmonised methodology for conducting a data protection impact assessment under the GDPR, and risk assessments under the future AI Act. Rather than dictating a particular type of explainability outcome, a methodology serves as a set of common steps to arrive at a solution that can then be defended as providing meaningful explanations of AI outputs to most users, under a best effort, risk-based, approach. We outline below what this methodology would look like.

## 5.1 A "Meaningfulness" Matrix

The discussion above has revealed two main variables that affect meaningfulness:

(a) the five purposes that explanations are intended to serve: controlling system performance, evaluating individual results, empowerment/redressing information asymmetries, contestability, and public algorithm transparency, and

(b) the four kinds of potential recipients of the explanations: data science teams, human operators of the system, persons affected by algorithmic decisions, and judges/regulators/auditors.

With these two variables - objectives and audiences - we have the beginning of a framework for analysing XAI solutions in light of the purposes and intended audiences. The right solution must also take into account the operational context in which explanations are given, particularly the time and other cognitive constraints facing the human recipient of the explanation.

By providing a common analytical framework for achieving meaningful XAI, the European Commission could, through guidelines, pave the way toward more specific standardisation work covering particular operational contexts, such as facial recognition, fraud detection, or medical diagnosis. For example, the Commission could provide general guidelines on the analytical steps leading to a definition of what constitutes a "meaningful" explanation in a given context, similar to guidance on how to conduct a data protection impact assessment. Standardisation bodies could then refine these guidelines in particular operational contexts. By following these standardised frameworks, AI developers could propose default XAI settings that emerge from the framework, justifying their XAI choices based on user-centric design and testing methods. These default settings proposed by the system provider would naturally need to be evaluated by the user of the system (such as a bank), who is ultimately responsible for the system's operational environment. The user may wish to conduct their own testing and decide to change the default settings to better suit the operational environment and particular user needs.

Following the harmonised methodology would not guarantee that a given explainability solution is always legally acceptable. Courts and regulatory authorities would have the final say on what XAI solution satisfies legislative and fundamental rights requirements in a given context. However, following the standardised methodology would permit system providers and users to demonstrate their compliance with obligations to provide effective and meaningful explanations using a risk-based, accountability, approach.

Importantly, the technical and user documentation associated with high-risk AI applications under the AI Act should present the operational choices and trade-offs that were made for defining meaningful local explanations, and why certain options may have been discarded by the provider or user in the course of user-centric design methodology. As we have seen from the GDPR, some design choices made on how to present user information can either be interpreted as enhancing understandability for the user, or on the contrary be criticised as hiding important information from the user in violation of law. This puts enterprises in a difficult position, leading often to the solution of providing too much information at the expense of understandability. By documenting choices made to enhance the understandability of individual explanations, design choices can more easily be defended under an accountability and risk-based approach.

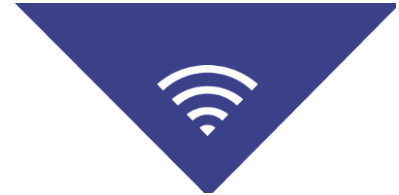

## 5.1. Examples of a Matrix for XAI

**Example 1:** Let us take the example of an algorithmic system deployed by a social media site to detect terrorist content online.[101] The system will generate an alert of possible terrorist content that is then acted upon by a human operator. In this case, the rows in the table below show the different audiences of explainability, and the columns show the different purposes of explainability. A checkmark shows where explainability is important for the audience and purpose, and whether "global" or "local" explainability is needed, or both.

| | | PURPOSES OF XAI | | | | |
|---|---|---|---|---|---|---|
| | | Controlling system performance | Evaluating individual results | Empowerment/ redressing information asymmetries | Contestability | Public admin transparency |
| AUDIENCES | Data science team | ✅ (global and local) | | | | |
| | Human operator | ✅ (global and local) | ✅ (local) | | | |
| | Person affected | | | ✅ (global) | ✅ (global and local) | |
| | Regulator /judge | ✅ (global and local) | ✅ (local) | | ✅ (global and local) | |

For each box with a checkmark, a human-centric design approach would help determine whether a given explanation method contributes (or not) to the identified purpose, given the type of audience and the operational context.

---

[101] Supra footnote 57.

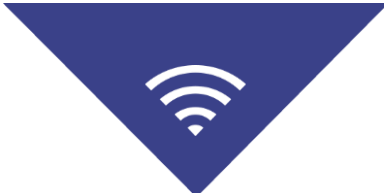

**Example 2:** The matrix for an algorithm deployed by police authorities to detect terrorism risks[102] would look similar, except that the need for public administration transparency would also come into play.

| | | PURPOSES OF XAI | | | | |
|---|---|---|---|---|---|---|
| | | Controlling system performance | Evaluating individual results | Empowerment/ redressing information asymmetries | Contestability | Public admin transparency |
| AUDIENCES | Data science team | ✅ (global and local) | | | | |
| | Human operator | ✅ (global and local) | ✅ (local) | | | |
| | Citizen affected | | | ✅ (global) | ✅ (global and local) | ✅ (global and local) |
| | Regulator /judge | ✅ (global and local) | ✅ (local) | | ✅ (global and local) | ✅ (global and local) |

---

[102] Such as the system examined by the CJEU in *La Ligue des Droits Humains*, op. cit.

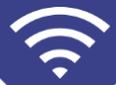

# 6. RECOMMENDATIONS FOR THE PROPOSED AI ACT

XAI covers many concepts and raises complex definitional questions. Dividing XAI into "global" and "local" explainability makes classification easier, even though it hides many nuances explored in the XAI literature. Global explainability refers to conveying an understanding of the system as a whole. Local explainability refers to conveying an understanding of a particular algorithmic result. Local explainability is not dealt with directly by the proposed European AI Act. The proposed AI Act requires that information about the system permit the persons in charge of human oversight to correctly interpret algorithmic results, and this requirement will in fact require local explanations, even though "local" XAI is never mentioned as such in the text. The AI Act contains no guidance on how the local explainability should be attained, and what is meant by "meaningful" explanations.

The absence of any specifications for local explainability in the AI Act is all the more unfortunate because many other EU texts require forms of local explainability. The AI Act would be an opportunity to create a harmonised approach to XAI that applies across different EU legislation.

Our suggested approach would be for the AI Act to mention the need for local explainability to be taken into account in the risk analysis and risk management system of the AI provider and adapted as necessary by the user of the AI system. The AI Act would then empower the Commission to develop guidelines on when and why AI explanations may be required, and how those explanations, whether global or local, should be "meaningful" in light of the purpose of the explanation and its audience. These guidelines, which might be inspired by the Commission's guidelines for the P2B Regulation, would propose a methodological approach to reaching "meaningful" explanations in a given context, using the matrix approach suggested above.

The guidelines could potentially go beyond the AI Act, and mention explainability requirements that emerge from other sources of EU law, including the case law of the CJEU. The guidelines would propose a harmonised methodology for approaching meaningful explanations in these different EU legal texts, recommending user-centric design methodology as one way of measuring whether a given explanation is "meaningful" in light of its audience and purpose.

These guidelines would not guarantee that any given solution developed by a provider or user of an AI system achieves a perfect approach to explainability in a given situation, but the methodology would show that the provider or user of the system acted diligently in developing an explainability approach that it believes provides the most meaningful XAI for the most people given the context, in the spirit of accountability.

Asking the Commission to develop guidelines for a harmonised approach to different explainability requirements will raise its own challenges. Different stakeholders were involved in each legal text, and the language used in each text is different. Nevertheless, the EU Charter applies to all, and as we have seen[103] the CJEU has imposed explainability in certain contexts with reference to the Charter. The Charter could therefore underpin each approach to XAI appearing in different legal texts and provide a foundation for seeking a harmonised methodology.

---

[103] Supra, Section 3.2.

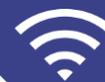

Another complementary approach would be for European standardisation bodies to develop the framework for meaningful explanations in the context of harmonised standards. This appears to be the route preferred by the Commission, which issued on May 22, 2023, a request to the European standardisation bodies CEN and CENELEC to develop specifications on the design and development of solutions that ensure transparency of the operation of AI systems to enable users to understand the systems' output and use it appropriately.[104]

Another complementary angle would be for the European Centre for Algorithmic Transparency, established by the Commission to deal with algorithmic transparency in the context of the DSA, to develop a methodology of the kind we propose here.

Regardless of the method, we have identified two objectives that the Commission should pursue:

- harmonisation of multiple XAI requirements appearing in different EU texts and case law;

- encouragement of user-centric design methodology to achieve "meaningful" explanations.

---

[104] Commission Implementing Decision of 22 May 2023 on a standardisation request to the European Committee for Standardisation and the European Committee for Electrotechnical Standardisation in support of Union policy on artificial intelligence, C(2023) 3215 final.

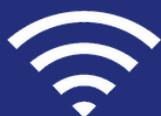

cerre · Centre on Regulation in Europe